\documentclass[prl,showpacs,amsmath,12pt]{revtex4}

\begin{document}

\title{$R$-function Related to Entanglement of Formation}
\author{Shao-Ming Fei$^{1,2}$}
\author{Xianqing Li-Jost$^{2}$}
\affiliation{$^1$Department of Mathematics, Capital Normal University, Beijing 100037,
China\\
$^2$Max Planck Institute for Mathematics in the Sciences,
D-04103 Leipzig, Germany}

\begin{abstract}
By investigating the convex property of the function $R$, appeared in
computing the entanglement of formation for isotropic states in
Phys. Rev. Lett. \textbf{85}, 2625 (2000), and
a tight lower bound of entanglement of formation
for arbitrary bipartite mixed states in
Phys. Rev. Lett. {\textbf{95}}, 210501 (2005), we show analytically
that the very nice results in these papers are valid not only for
dimensions 2 and 3 but any dimensions.
\end{abstract}

\maketitle

Quantum entangled states are playing fundamental roles in
quantum information processing \cite{nielsen}. The entanglement of formation (EOF)
is a most important measure in quantifying the degree of entanglement
\cite{BDSW,Horo-Bruss-Plenioreviews}.
Considerable efforts have been spent on deriving EOF or its lower bound
through analytical and numerical approaches (\cite{Terhal-Voll2000,caf}
and references therein). In \cite{Terhal-Voll2000}
the EOF for isotropic states are presented.
It is shown that for $F\geq 1/d$, the EOF for isotropic
states $\rho _{F}$ is $E(\rho _{F})=co\big(R(F)\big)$, where
\textquotedblleft co($R$)" stands for the convex hull,
the largest convex function that is bounded above by the function $R$.
An explicit expression of $co\big(R(F)\big)$ has been derived for dimensions
$d=2,3$, and its general form is conjectured for arbitrary $d$.
While in \cite{caf} a tight lower bound of EOF
for arbitrary bipartite mixed states is given, by using the property of $R$ function
derived in \cite{Terhal-Voll2000}.
For any bipartite $m\otimes n$ $(m\leq n)$ mixed quantum state $\rho$, the EOF
$E(\rho)$ satisfies $E(\rho )\geq co\big(R(\Lambda)\big)$.
In both cases the results depend on the convex property of the function $R$.
These results are correct when the second derivative of $R$ has one zero point,
which has been shown to be true if the dimension is 2 or 3.

The $R(\Lambda )$-function ($\Lambda$ corresponds to $d\, F$
in $R(F)$ \cite{Terhal-Voll2000}) has the form
\begin{equation}\label{R}
R(\Lambda )=H_{2}\big(\gamma (\Lambda )\big)+\big(1-\gamma (\Lambda )\big)
\log _{2}(m-1),
\end{equation}
where
\begin{equation}
\gamma (\Lambda )=\frac{1}{m^{2}}\left(\sqrt{\Lambda }+\sqrt{(m-1)(m-\Lambda )}\right)^{2}
\label{gamma}
\end{equation}
and $H_{2}(.)$ is the standard binary entropy function,
$\Lambda\in [1,m]$. For $m\geq 4$, it is conjectured that the
second derivative of $R$ w.r.t. $\Lambda$ has still only one zero
point, by numerically calculating the function $R$ for a given
$m$. In deed for $m=4$, one can easily see this is true by
plotting $R(\Lambda)$. In the following we show analytically that
the results in both papers valid for arbitrary dimensions $m\geq
5$.

For simplicity we replace $\log_{2}$ in (\ref{R}) by the natural $\log$.
Without confusion we still use the notion $R(\Lambda)$ below, which, in fact,
differs a positive factor $\log_2 e$ from the $R(\Lambda)$ above.
We first prove that there is one and only one point $\Lambda_0$
between $1$ and $m-1$ such that $R^{\prime\prime}(\Lambda_0)=0$
for $m\geq 5$. Then we further show that there is no more zero points
for $R^{\prime\prime}(\Lambda)$ between $m-1$ and $m$.
By direct calculation we have the second derivative of $R$ w.r.t $\Lambda$
\begin{equation}\label{ddR}
R^{\prime\prime}(\Lambda)=\gamma^{\prime\prime}(\Lambda)
\log\frac{1-\gamma(\Lambda)}{(m-1)\gamma(\Lambda)}-\frac{1}{\Lambda(m-\Lambda)},
\end{equation}
where
\begin{equation}\label{ddg}
\gamma^{\prime\prime}(\Lambda)=-\frac{\sqrt{m-1}}{2}(\Lambda (m-\Lambda))^{-3/2}.
\end{equation}
From which we get $R^{\prime\prime}(1)=\text{Lim}_{\epsilon\to 0}
R^{\prime\prime}(1+\epsilon)=+\infty$. On the other hand,
$$
R^{\prime\prime}(m-1)=-\frac{1}{m-1}\left(\log\frac{m-2}{2(m-1)} +1\right),
$$
which is less than $0$ for $m\geq 5$. Therefore for $m\geq 5$ there exits
$\Lambda_0\in (1,m-1)$ such that $R^{\prime\prime}(\Lambda_0)=0$. From
(\ref{ddR}) and (\ref{ddg}) $\Lambda_0$ is the solution of $g(\Lambda)=f(\Lambda)$, where
$$
g(\Lambda)=\log\frac{1-\gamma(\Lambda)}{(m-1)\gamma(\Lambda)},~~~
f(\Lambda)=-2\sqrt{\frac{\Lambda(m-\Lambda)}{m-1}}.
$$
As $g^\prime(\Lambda) > 0$, $g(\Lambda)$ is a monotonically increasing function
taking values from $g(1) \to -\infty$ to
$$
g(m-1)=2\log\frac{m-2}{2(m-1)}>-2.
$$
While $f(1)=f(m-1)=-2$, $f^{\prime\prime}(\Lambda) > 0$, i.e. $f$ is convex.
Therefore there is one and only one solution $\Lambda_0$ to the equation
$g(\Lambda)=f(\Lambda)$ for $\Lambda\in (1,m-1)$.

We now show that there are no more solutions to $R^{\prime\prime}(\Lambda)=0$
for $\Lambda\in (m-1,m)$, i.e. $R^{\prime\prime}(m-1+\delta)\neq 0$, $\forall\,
\delta\in (0,1)$. From (\ref{gamma}), (\ref{ddR}) and (\ref{ddg}) this is equivalent to show
$F(\delta)\equiv \frac{1}{2}B(\delta)\log A(\delta)\neq -1$, where,
$$B(\delta)= \sqrt{\frac{m-1}{(m-1+\delta)(1-\delta)}},~
A(\delta)= \frac{(mC(\delta))^2-1}{m-1},$$
and $C(\delta)= \big(\sqrt{m-1+\delta}+\sqrt{(m-1)(1-\delta)}\big)^{-1}$.
It is straight forward to verify that $A(0)>0$.
As the derivative of $C(\delta)$ w.r.t $\delta$, $C^\prime(\delta)>0$, we have
$A^\prime(\delta) >0$. Hence $\log A(\delta)$
increases as $\delta$ increases. Similarly, as
the derivative of $(m-1)/\big((m-1+\delta)(1-\delta)\big)$ w.r.t $\delta$ is positive,
$B(\delta)$ also increases as $\delta$ increases.
Therefore $F(\delta)$ is an increasing function of $\delta$.
Moreover $F(0)=\log(m-2)/\big(2(m-1)\big)\geq \log3/8>-1$. We have
$F(\delta)\geq F(0) > -1$, $\forall\,\delta\in (0,1)$ and $m\geq 5$.
Thus $R^{\prime\prime}(\Lambda)=0$ has no solutions for $\Lambda\in (m-1,m)$.

We have shown that the $R$-function has only one reflection point
for any dimensions. Therefore the construction of
the largest convex function that is bounded above by the $R$-function in
\cite{Terhal-Voll2000} is correct. Either the EOF for isotropic states
in \cite{Terhal-Voll2000} and the tight lower bound of EOF in \cite{caf}
are valid for arbitrary dimensions.

\vspace{0.5truecm}
\noindent {\bf Acknowledgments}\, We thank Q. Chen and Y.H. Yang for very
helpful discussions. S.M. Fei gratefully acknowledges the warm hospitality
of Dept. Phys., National University of Singapore, where the work is finished.
The work is partially supported by NKBRPC (2004CB318000).


\begin{thebibliography}{99}
\bibitem{nielsen} M.A. Nielsen and I.L. Chuang, Quantum Computation and
Quantum Information, Cambridge University Press, Cambridge, 2000.

\bibitem{BDSW} C.H. Bennett, D.P. DiVincenzo, J.A. Smolin, and W.K.
Wootters, Phys. Rev. A \textbf{54}, 3824 (1996).

\bibitem{Horo-Bruss-Plenioreviews} M. Horodecki, Quant. Inf. Comp. \textbf{1}%
, 3 (2001); D. Bru\ss , J. Math. Phys. \textbf{43}, 4237 (2002); M.B. Plenio
and S. Virmani, quant-ph/0504163.

\bibitem{Terhal-Voll2000} B.M. Terhal and K.G.H. Vollbrecht, Phys. Rev.
Lett. \textbf{85}, 2625 (2000).

\bibitem{caf} K. Chen, S. Albeverio and S.M. Fei, Phys. Rev.
Lett. \textbf{95}, 210501 (2005).
\end{thebibliography}
\end{document}